\documentclass[12pt]{article}
\usepackage{graphicx}
\usepackage{amssymb,amsmath,amsfonts,palatino,amsthm}
\usepackage{amssymb}
\usepackage{epstopdf}
\newcommand{\f}{\begin{equation}}
\newcommand{\ff}{\end{equation}}

\begin{document}
\title{The quantum mechanics of the present}
\author{Lee Smolin${}^a$ and Clelia Verde${}^b$
\\
\\
${}^a$ Perimeter Institute for Theoretical Physics,\\
31 Caroline Street North, Waterloo, Ontario N2J 2Y5, Canada\\
and\\
Department of Physics and Astronomy, University of Waterloo\\
and\\
Department of Philosophy, University of Toronto\\
\\
\\
${}^b$Corso Concordia,   20129 - Milano}
\date{\today}
\maketitle



\begin{abstract}
We propose a reformulation of quantum mechanics
in which the distinction between definite and indefinite becomes
the fundamental primitive.  

Inspired by suggestions of Heisenberg, Schrodinger
and Dyson that the past can't be described in terms of 
wavefunctions and operators, 
so that the uncertainty principle does not apply to past events,
we propose that the distinction between past, present and future
is derivative of the fundamental distinction between indefinite
and definite.  

We then outline a novel form of presentism based on a phenomonology
of events, where an event is defined as an instance of transition
between indefinite and definite.  Neither the past nor the future fully
exist, but for different reasons.  We finally suggest reformulating physics
in terms of a new class of time coordinates in which the present time
of a future event measures a countdown to the present moment in which that
event will happen.

\end{abstract}
\newpage
\tableofcontents

\section{Introduction}

The idea we will discuss here has arisin from time to time since the invention of quantum mechanics.  Here are the earliest quotes we've found so far, courtesy of 
Jeremy Bernstein\cite{JB}.

\begin{quotation}

{\it  ``This formulation makes it clear that the uncertainty relation does not refer
to the past."}

-Werner Heisenberg\cite{Heisenberg}.

\end{quotation}

\begin{quotation}
{\it "For eternally and always there is only now, 
one and the same now.  The present
is the only thing that has no end."}

-Erwin Schrodinger\cite{Schrod}

\end{quotation}

\begin{quotation}

{\it I deduce two general conclusions from these thought-experiments. First,
statements about the past cannot in general be made in quantum-mechanical
language. We can describe a uranium nucleus by a wave-function including an
outgoing alpha-particle wave which determines the probability that the nucleus
will decay tomorrow. But we cannot describe by means of a wave-function the
statement, ``This nucleus decayed yesterday at 9 a.m. Greenwich time''. As a
general rule, knowledge about the past can only be expressed in classical terms.
My second general conclusion is that the ``role of the observer'' in quantum
mechanics is solely to make the distinction between past and future. The role of
the observer is not to cause an abrupt ``reduction of the wave-packet'', with the
state of the system jumping discontinuously at the instant when it is observed.
This picture of the observer interrupting the course of natural events is
unnecessary and misleading. What really happens is that the quantummechanical description of an event ceases to be meaningful as the observer
changes the point of reference from before the event to after it. We do not need a
human observer to make quantum mechanics work. All we need is a point of
reference, to separate past from future, to separate what has happened from
what may happen, to separate facts from probabilities. }

-Freeman Dyson\cite{Freeman}

\end{quotation}

There seemed to be something correct in these statements, but  they also leave
us puzzled.   

This view seems to treat the future, present and past differently from each other - 
but it is also  not
{\it presentism} - which posits that the future and past don't exist; only the present
is real and exists. For if the past is described by classical mechanics - certainly
it may exist.  And if the future is describable by the quantum wavefunction -
certainly then it represents something that may come to exist in the present. 
This seems to be true whichever 
interpretation of quantum mechanics you espouse.  

There is a view, which we may call {\it block-presentism}, according to which past events, present events and future events all exist, at all times, but their status changes.  It seems as if future, present and past are just labels that can be pinned on otherwise timeless events, which has no effect on their other properties.

Then there are hybrid formulations such as the ``{\it growing block universe"}
according to which the past and present are real, but the future is not.  But in what sense does the past exist - if there is no way for us to alter it?  At least not if one is also
aspiring to be a relationalist - for whom anything that exists does so as part of an evolving network of relations and interactions.

Taking the passage of time seriously must be more than to draw a thin green line
on the block universe to show which slice is presently now.  

We find the following analogy helpful:  perhaps an observer in the present moment feels no passage of time, the same way a free falling observer feels no gravity.   So we began wondering if we could transform  ourselves into the rest frame of a present moment observer, who would see themselves always at the origin of the time coordinate.



From neuroscience we know that it is possible that 
  -  with respect to the    universe's time  - we live in the past and there is no present There is only the past and the future.

It is also  possible to define the present as the moment our consciousness becomes aware of the past.     

This paper is then an attempt to invent a setting within which these thoughts of Heisenberg, Schrodinger and Dyson may make sense.

\section{Constructions of space and time}

Einstein taught us that our representations of events far from observers are constructions, that is hypotheses as to distant parts of the world \cite{AE1905}.

We tiny, local observers become aware of distant events only from the light, gravitational radiation and other radiation they send us. We may also
probe the universe by sending out light signals and watching for their return as in radar.  The times that signals arrive,
leave us or return back can be labeled by readings of a clock (by that we mean a physical clock, local to us and weakly coupled so we read it without destroying it).  And that's it. From that meagre data alone we construct a map of the universe, into which we fit our knowledge of its history.


This philosophy is also extended to our description of the microscopic world,
as it is seen clearly in operational formulations of quantum theory, in which the complexities of the atomic world are broken down into the three stages of
preparation, evolution and observations \cite{Lucien}. These stages can also be seen as instances of probing and receiving light signals, at times recorded on a 
laboratory clock.

In this paper we ask what happens to these constructions when we consider
time from a different point of view. We suppose to make a construction of the world 
that, rather than employing the usual distinction between past, present and future, is based on a single fundamental and objective distinction, the indefinite and  the definite.  

To make this distinction absolute we need to fix a basis or, what amounts to the same thing, choose a configuration space.  Then a state is definite if it has a 
definite value of the corresponding observable.     Otherwise it is a superposition of two or more eigenstates and we say it is an indefinite state.   We regard this as a good
thing, even if it breaks the symmetry of Hilbert space.

We will argue elsewhere\cite{elsewhere} that the distinctions philosophers take to hold between past, present and future can be reduced to and explained by the distinction between indefinite and definite. 

On the other hand,  in quantum theory, the distinction between definite and indefinite appears to underlie the distinction between
amplitude and probability.  

We can say that when we observe something we do so as spectators to the transition between indefinite and definite.  In this way we aspire to construct a compact view of the world compatible  with quantum theory and relativity.    

\section{A phenomonology of present events}

Chasing down the thoughts of Heisenberg, Schrodinger and Dyson
leads us to what we might call a phenomenology of time.   This can be considered
an interpretation of quantum theory - or else a substitute for one.  It is
a phenomenology in the sense that it organizes how time appears to prevade
our experience of the world.

We next state the central points of this phenomeology.

\subsection{The definite and the  indefinite}

\begin{itemize}

\item{} Everything that is real is so in the present moment, which is one of
a succession of moments.


\item{} This is a phenomenology of present events. Nothing exists or persists, things only happen.  

\item{}   What we mean by {\it becoming} or {\it to happen} is for something
indefinite to become definite.  
This is what we call  {\it an event}.     Events are real.

\item{}What becomes definite at any event is not arbitrary.  Events
happen for a reason, which comes from causally prior events.


\item{} Events exist for a definite duration and then give rise to subsequent events, which make up subsequent present moments.   This is how a world based only on the distinction between the indefinite and the definite may define also a causal structure.

\item{}The quantities that become definite at any event are its endowments,
which are
passed to each from preceding events and become definite on reception.

\item{}These endowments include energy and momentum.  Each event
is brought to happen by the passage of the endowments, from their immediate
predesescors to them.
These instances of passing on, define the causal relations, which are definite.

\item{} Only the becoming - the transition from indefinite to definite - is real. To  exist is to trace a transitory event, and  these transitions are what we call the now - the present.  

\item{}One might try to keep the past as real,  with the present as the future
oriented boundary of the spacetime.   But this would be highly wasteful given that the past cannot have any further influence on the present.

\item{}  A set of events may be co-present, making up the same present moment.
A present moment may be thick, i.e. contain or overlap with pairs of events that are causally related.

\item{}The direction from indefinite to definite gives the universe an arrow of 
time.



\item{}The quantities that become definite at an event are called the
view of the event.  The views are real.






\item{}Once an event that was indefinite,  becomes definite, it cannot immediately 
reverse and become indefinite again.  Once an event happens, it cannot be made to unhappen.



\subsection{The past}
\item{} Events  have no past. The influence of the past is already present in the
present.

\item{}We can change neither the past nor the future, but for different reasons.
We cannot change the past because it has already become definite. Our only power is to influence what events, which yesterday were indefinite,
become definite and - in some cases - what choices were 
made.That is, we
can only act in the present to influence the process by which the formerly indefinite becomes definite

\item{}We can only change the present and this consists at best of resolving the indefiniteness with which the future confronts us.


 \item{} Traces of these causal partial orders partly constrain the initial conditions for the becoming of the new events.    In some cases these  are captured in patterns within a single present moment,  to become partial and 
 incomplete "records of the past".  These give evidence that there may have been previous present moments.     

\item{}In some cases these "records" may be detailed enough to allow us to construct a hypothetical science of the dynamics of past moments.  We can check records of experiments that were carried out in prior present moments with theoretical predictions derived in our present moment.  This is ultimately a exercise in re-constitutional fiction.  

\subsection{The future}

\item{The future does not yet exist.} The future  will consist of events 
at which present indefinites will turn into definites.

\item{}The future is underdetermined.  But we can imagine various paths
it might evolve from our present moment, and make bets, or -what is the same - assign probabilities to them.

\item{}The indefinite is also called the future. This is because being indefinite it can at any time become definite - in an event which is definite and real. If it does, it may influence the present moments to come.  

\item{}The quantum state is nothing but an expression of what we can best
forecast or bet about the future, taking fully into account both what is indefinite and definite at this present moment. 

\item{}Present moments exist for a finite duration and give  rise to successor present moments.  
But, when they do, that previous present moment no longer exists, and the the next present moment is made up of novel events.

\item{} We can call events that we speculate may follow the present events, possible  future events.  But they are not part of reality, because reality is limited to the present moment.   We can speak of possible future moments, so long as we remember that they are not real, they are speculations about what may become real later.

\item{} The future is the name we give to speculation as to which events will happen, later in reality, which of course means that they are speculations about how indefinites will turn definite, in future present moments.









\item{}The world recreates itself in every moment, as indefinites flash into
momentary definites, after which they are nothing. Everything we see around us exists or did just exist, but was gone in the blink of an eye.

\end{itemize}

\subsection{Causality without determinism}

\begin{itemize}

\item{} The world is causal.  All events have causes and all are
causes of other events.
The world is generally not deterministic. Pure quantum theory
is deterministic in that the evolution of the wavefunction - the Schrodinger equation
is deterministic.  If you input a definite initial state it will output
a definite finite state. That is, it is the indefinite that evolves deterministically.

\item{} The transition from indefinite to definite is non-deterministic, and only given by a stochastic rule.

\item{} This is how it must be if Leibniz's Principle of the Identity of the
Indiscernible is to be satisfied.   For if all the inputs into an event are identical, the outputs must be diverse and not predicted.  How much diversity?  Enough that each event is uniquely distinguishable.





\item{} Time's activity is in creating each new event - i.e. by repeatedly bringing an
indefinite to an unpredictable but definite state.

\item{} So quantum dynamics is perfectly paradoxical:  
the indefinite evolves deterministically,  while
the definite becomes so non-deterministically, b?ut causally.



\section{The quantum mechanics of definite and indefinite}

We have presented a radically different ontology of time than is usually used in physics
\footnote{For example, some think it preferable to formulate quantum mechanics in time reversal symmetric form
\cite{APC}.}, 
which is, compared to the block universe picture, much sparser. We eliminate both past and future from the description of reality and keep only an ever-changing band of  present events. The past is represented in the present in records, fossils and the like. These give us reason to believe there were previous moments  But there is no reason to believe those moments still "co-exist" with us. What would that add? 


What holds up the old house where we sit comfortably writing is not the weight or any aspect of the past, but literally its continual recreation, moment to moment.  

The future also does not exist presently as indeed it can have no causal effect on us
(any such influence would be retrocausal: from the future to the past).

Saying that the future does not now exist is not saying that anything is possible.  The point of having laws of physics is that they restrict what can happen in the future.

The present evidence for laws comes from the past,  by comparison of records of past
completed observations and experiments and past predictions for the results of those
experiments.

Using this knowledge we can distinguish future evolutions that are possible from those that are not. This does not mean we give them different ontological status.

Now the most important question we have to answer is whether the change of ontology we are suggesting implies we must alter the practice of quantum
mechanics. This is a slightly subtle  question because the different formulations
of quantum mechanics do disagree with each other when it comes to the treatment
of macroscopic quantum superpositions.

The first thing to say is that ours is an 
events 
ontology\footnote{And is thus compatible with the approaches to quantum gravity and quantum foundations developed in \cite{related}-\cite{related7}.}.   

We don't posit that the world is composed of "states" and we mention them only to specify certain ensembles.  So if we haven't checked the life of the  cat for a few seconds - which will be
an event when we do so, it is not useful to imagine that there is a cat-state.

Let us look again at the quotes from Heisenberg and Dyson. We believe it is clear that what Dyson has in mind is a revision of the Copenhagen formulation in which
there remains the two complementary realms, quantum and classical, but the
boundary between them is moved from microscopic versus macroscopic to instead
divide the future from the past.  

Although Heisenberg's quote is much shorter it seems reasonable to imagine he meant something similar.  

Of course, just like in Copenhagen, we are allowed to use a classical
description of events in the past.  We just must be aware that we
do not gain any increase in predictive power.   For example, we can measure the linear polarization of one of two entangled photons in the EPR set up, and the circular polarization of the other.   Making use of the conservation laws one can then assign
both kinds of polarization to each photon.  But because that requires an exchange of classical information, you cannot use
the results to send signals faster than light.   Realists may use this to argue that quantum mechanics is incomplete.

To the extent that we are following Dyson in repositioning the boundary in Copenhagen we can call this the {\it  temporal Copenhagen interpretation.}.
And it looks pretty likely we can make an argument that  as far as
the practice of quantum mechanics is concerned,  temporal Copenhagen
is equivalent to Copenhagen.

This is good because no experiments are known, presently,  whose
results are inconsistent with the Copenhagen formulation.

Is our proposal consistent with the other approaches to quantum
foundations?


We emphasized that our first principle is that the fundamental distinction in nature is
between indefinite and definite. Is there an equally basic distinction in the quantum theory that mirrors this distinction?  

The correspondence is:

\begin{itemize}

\item{}if a process or observable is {\it indefinite}  the amplitudes for the different possibilities are {\it summed over in the path integral}, before the absolute values squared is taken to give probabilities;

\item{}the amplitudes for definite processes are not summed over - rather we form  the probability for each definite process by directly taking the absolute value squared. Often this is all you have to do because there is just one definite history.

\end{itemize}

This correspondence is fundamental to quantum theory, but it cannot be the whole story, because it doesn't dictate when the two rules are applied.   When are the amplitudes of the  quantum histories summed over and superposed and when is there
a real event, where only probabilities are summed? And are these consequences
of external interactions as in von Neumann's two rules for measurements assert?

Our proposal fills in this gap in a very simple way, that uses no additional
structure, because in our proposal the distinction between definite and indefinite is a primitive that does not require further explanation. When making  a prediction about the future, sum over amplitudes and then
take the absolute value squared of the sum to get probabilities.

When processing records from the past, take the absolute value squared of the amplitude for each history and then add the resulting probabilities.


\section{The frame of reference for an observer in a present moment}

We have given a  perspective on the past and future in terms of indefinite
becoming definite.   To realize the implications of this change we introduce a new  time coordinate and indeed a new kind of reference frame, that parameterizes what an observer in a present moment would experience.  This is a family of time coordinates, one for every present moment, in the succession of such moments, which gives a special indication to the present present moment.


Because of {\it the flow of time} the present moment keeps becoming, creating new events which become present just as the formerly present become past.  Just as one has to keep translating in space to keep the inertial observer at the origin of the spatial coordinates (always at $x^i =0$)   we have to continually translate the present moment observer into the future to keep them at the temporal origin $\tau =0$.

We need to introduce  two classes of time coordinates:

\begin{itemize}

\item{} {\it clock times} $T$,  are readings on a physical inertial clock, which is not part of the system being studied.  

There are an infinite number of clock times, which in the absence of gravity,
are related to each other by Poincare transformations. They all express
the principle of inertia, namely that they can each declare themselves to be at
rest, and measure all motion relative to them.

\item{}{\it present moment relative times}, indicated by $\tau$.  

An observer in each of the succession of present moments can construct 
one of these. Each one is defined so  that $\tau$ in  its  present  moment vanishes.

\end{itemize}

Let us assume  that we have chosen an inertial clock, $T$.
The reading of that clock in a particular present moment
distinguishes that moment and is called $T_0$.
We can then define:
\f
\tau = T_0-  T
\label{tau}
\ff

The reason we need two kinds of time coordinates is that quantum physics is causal but not deterministic. The causality is represented by the present moment coordinates which build the current previous moments on the previous ones.  This is the true action of dynamics in quantum physics.  The basic idea of using $\tau$ to measure time is that, as we evolve through
a succession of present moments, $\tau$ always evolves so that is always clear which time is the present present moment: it is the one in which $\tau =0$.    The clock time connects the system to the rest of the world without imposing determinism.

We note that $\tau$ evolves in the opposite sense of $T$.  Suppose we
start out in a present moment $T_{_1} $ 
which is before $T_0$, i. e. the present
moment indicated by 
$T= T_0$ is then  in our future.  In terms of clock time, $T_0$ remains fixed in the  future,
and our clock increases, which gives a sense that  we are rushing towards that future (isn't that how it often feels?).  

When we use (\ref{tau}) to transform  to the $\tau$ coordinate,
it is like transforming to a moving frame, to capture the idea that an inertial  observers feel no motion. The present moment coordinate starts in the past of that present moment, and sits
stationary at $\tau=0$, while the future rushes towards it.

We are to a certain extent - but not completely - in a circumstance analogous to that of a local inertial observer.  They construct an hypothesis as to the world far from them
using only the information they have locally, which comes from the light signals they send out (probes) and receive (views) at times measured by their local clock.

\end{itemize}

It is instructive to imagine what it would be like to live in a world that ran on present moment time; the present time would always be $\tau=0$.    We wouldn't need
clocks to tell us that the present moment would always be NOW.

But the times assigned to future events would not be static, as they would always be counting down to the particular present moment when they would "happen" and emerge from the indefinite to the definite.

Rather than living by a clock which always ran, to orient ourselves to a static calendar,
the calendar would indicate a complexity of countdowns, each evolving towards its moment.   




\section{Closing remarks}

Our proposal is not subjective or relational. Some people who  embrace  relational approaches to quantum mechanics, will regard  our existence, no matter how definite for us, as indefinite to them. We do not make any such claim.  We propose that definite is an objective and universal status, which is why our proposal
requires a choice of configuration space, or basis.

There are several ways our proposal can 
be read\footnote{There are other contexts in which retrocausality
plays a role in proposed reformulations of 
quantum theory\cite{retro1}-\cite{retro7}.  }.    
One can
read definite and indefinite as epistemological terms --  stating the status
of the knowledge we have of a physical system.  Or one can give the distinction an ontological reading. Or we can,
following Aristotle,  think about the distinction between the definite and the indefinite
a distinction between ways of being - a potential way of being 
contrasted with an actual way 
 of being\cite{KK}.      
 
So, finally, we define the present, past and future in terms of the transition from
indefinite to definite as follows.  Reality, which is to say the momentary present, consists of nothing but flashes of
transitions from indefinites to definites, which persist for an instant in their definitness and then are gone.  This has happened before, but once something is definite, its done its job and is gone.  There is no past, because once an event has done
its job it plays no further role.  The future does not concretely
exist, it contains indefinite quantities that have yet to
be rendered definite - and hence existing.
 
Indefiniteness can persists as such, but without existing.   This is the fate of the
future. By a resolution, an embrace of a definite, it briefly has existence, then is gone.

There is nothing under our feet, the past is gone.  


\section*{Acknowlegements}

We are grateful to Giovanni Amelino-Camelia, 
 Herbert Bernstein, Saint Clair Cemin,  Marina Cortes, Stuart Kauffman,  Joao Magueijo, Roberto Mangabeira Unger, Carlo Rovelli, and especially to  Jaron Lanier for critical questions and encouragement.

This research was supported in part by Perimeter Institute for Theoretical Physics. Research at Perimeter Institute is supported by the Government of Canada through Industry Canada and by the Province of Ontario through the Ministry of Research and Innovation. This research was also partly supported by grants from NSERC, FQXi and the John Templeton Foundation.


\end{document}